# Thermodynamic parameters of atomically thin superconductors derived from the upper critical field

Evgeny F. Talantsev

M.N. Mikheev Institute of Metal Physics, Ural Branch, Russian Academy of Sciences, 18, S. Kovalevskoy St., Ekaterinburg, 620108, Russia

## Abstract

The amplitude of the ground-state superconducting energy gap $\Delta(0)$ and relative jump in the electronic specific heat at the transition temperature $\frac{\Delta C}{\gamma T_c}$ are the primary fundamental parameters of any superconductor. Several well-established techniques are available for measuring these values in the bulk samples. However, a limited number of techniques can be used to measure these parameters in atomically thin superconductors. Here, we propose a new approach for extracting $\Delta(0)$ and $\frac{\Delta C}{\gamma T_c}$ in atomically thin superconductors by utilizing the upper critical field data from perpendicular, $B_{c2,\perp}(T)$ (when a magnetic field is applied perpendicular to the film surface), and parallel, $B_{c2,\parallel}(T)$ (when a magnetic field is applied in the direction parallel to the film surface), external field directions. The deduced parameters for few-layer-thick Al, Sn, NbSe$_2$, MoS$_2$, magic angle twisted trilayer graphene (MATTG), and WTe$_2$ are well-matched values expected for strong- and moderately strong-coupled electron-phonon-mediated superconductors. In many reports, the enhancement of $B_{c2,\parallel}(0)$ above the Pauli-Clogston-Chandrasekhar limiting field (i.e., magnetic field required to break the Cooper pair) in atomically thin superconductors has been explained based on the assumption of exotic pairing mechanisms, for instance, Ising-type pairing. Here, we explain the observed $B_{c2,\parallel}(0)$ enhancement based on the geometrical enhancement factor that originates from the sample geometry. This approach does not assume the existence of novel exotic pairing mechanisms in atomically thin superconductors.

**Thermodynamic parameters of atomically thin superconductors derived from the upper critical field**

## I. Introduction

Recently, Cao *et. al.* [1], and Zhou *et. al.* [2] measured the temperature-dependent upper critical field in twisted trilayer graphene (TTG) superlattices when an external magnetic field was applied parallel to the film surface, $B_{c2,\parallel}(T)$, and reported the extrapolated ground state upper critical field, $B_{c2,\parallel}(0)$, which was obtained from the fit of $B_{c2,\parallel}(T)$ data [1,2] to the equation:

$$B_{c2,\parallel}(T) = B_{c2,\parallel}(0) \times \sqrt{1 - \frac{T}{T_c}} \qquad (1)$$

where $T_c$ and $B_{c2,\parallel}(0)$ are free-fitting parameters, by several times exceeds [1,2] the Pauli-Clogston-Chandrasekhar limiting field, $B_{CPP}(0)$, for this material:

$$1.5 \times B_{CPP}(0) \lesssim B_{c2,\parallel}(0) \lesssim 3 \times B_{CPP}(0) \qquad (2)$$

where $B_{CPP}(0)$ was assumed to be [1,2]:

$$B_{CPP}(0) = 1.86 \times T_c. \qquad (3)$$

Similar observations have been reported for atomically thin $MoS_2$ [3-5], $NbSe_2$ [6-9], $MoTe_2$ [10,11], few-layer stanene [12], $PdTe_2$ [13], $TaS_2$ [7], ion-gated $SrTiO_3$ [14], and $WTe_2$ [15,16] (detailed reviews of recent studies in this field can be found elsewhere [17-22]).

Based on the observed violation of the Pauli-Clogston-Chandrasekhar limiting field (Eq. 2), and other research groups (see Ref. 1) proposed that magic-angle twisted trilayer graphene (MATTG) is a unique quantum matter material that can be considered as a material for next-generation quantum devices, including quantum computers.

Here, we should point out that the first report a many-fold increase in $B_{c2,\parallel}(T)$ vs. the decrease in film thickness, $d_{sc}$, (from 450 nm to 20 nm) was reported by Blumberg and Douglass [23,24] for Sn films in 1962. The reported increase in $B_{c2,\parallel}(T)$ was so high that the maximum available magnetic field of $B_{appl} = 0.3\ T \cong 10 \times B_c(0)$ (where $B_c(0)$ is the ground state thermodynamic critical field of bulk tin [24]) allowed the measured $B_{c2,\parallel}(T)$ dataset for 20 nm thick film [23] only within a close proximity to the transition temperature: $0.965 \leq \frac{T}{T_c} \leq 1.0$.

In the first study, where the extrapolated $B_{c2,\parallel}(0)$ exceeded the $B_{CPP}(0)$ (Eq. 3), was reported by Tedrow et al. in 1970 [25], who found that a 5-nm thick Al film exhibits the following:

$$B_{c2,\parallel}(0) = 4.84\ T > B_{CPP}(0) = 1.86 \times T_c = 4.56\ T. \qquad (4)$$

Later [26], the same group confirmed this result for several Al films with thicknesses $d_{sc} < 20\ nm$, and in 1982, the same group [27] reported a record high $B_{c2,\parallel}(0)$ in 4-nm thick Al films coated with 0.2 nm-thick Pt film:

$$B_{c2,\parallel}(T = 0.55\ K) = 9\ T > 4.5 \times T_c = 2.4 \times B_{CPP}(0). \qquad (5)$$

Considering that Tedrow et al. [25-27] utilized the 50% normal state resistance criterion to define $B_{c2}(T)$ in their films, while in some recent papers, a much less restrictive criterion (for instance, 75% of the normal state resistance [16]) was used, it can be concluded that the $B_{c2,\parallel}(0)$ enhancement observed in atomically thin Al films (Eq. 5 and Ref. 27) is identical to the enhancement reported for a various of van der Waals and electric-field-gated atomically-thin crystals recently [1-17].

This implies that the key parameter for the observed $B_{c2,\parallel}(0)$ enhancement in atomically thin superconductors is the film thickness, $d_{sc}$, because an identical effect is observed in pure elemental Al and Sn on the one hand, and in NbSe$_2$, MoTe$_2$, TTG, few-layer stanene, PdTe$_2$, TaS$_2$, electric-field-gated SrTiO$_3$, WTe$_2$, and MoS$_2$; on the other hand, it cannot be classified as exhibiting any similarity except being atomically thin.

Considering that proper equation for $B_{PCC}(0)$ [28] is:

$$B_{PCC}(0) = \frac{\Delta(0)}{\sqrt{g} \times \mu_B} \tag{6}$$

where $\Delta(0)$ is the ground-state superconducting energy gap amplitude, $\mu_B$ is the Bohr magneton, and $g$ is the Lande factor. Also, it should be stressed, that Eq. 6 has been derived for the case of infinitely large isotropic 3D superconductor. To the best of the authors knowledge detailed theoretical consideration of the angular dependence of $B_{PCC}(0, \theta)$ (where $\theta$ is an external field direction) for iso- and anisotropic superconductors with spatial confinement within several atomic layers has not been performed. Instead, in many reports on atomically thin superconductors, Eq. 6 is using to declare that because the observed upper critical field $B_{c2,||}(0)$ exceeds $B_{PCC}(0)$ (calculated by Eq. 6), atomically thin superconductors exhibit exotic type of charge carrier pairing.

By looking closer on Eq. 6, it can be concluded that Eq. 6 can be simplified into Eq. 3 only for clean *s*-wave weak-coupling superconductors that exhibit:

$$g = 2 \tag{7}$$

and

$$2\Delta(0) = 3.53 \times k_B T_c. \tag{8}$$

However, for many superconductors, even for *s*-wave lead, Eq. 8 is not satisfied:

$$2\Delta(0) = 4.5 \times k_B T_c \text{ (for lead [29])} \tag{9}$$

For pnictides, experimental data shows [30] that some of these superconductors exhibit a gap-to-transition temperature ratio as large as:

$$2\Delta(0) \gtrsim 9 \times k_B T_c. \tag{10}$$

Based on multiple statements [1,31,32] that MATBG and MATTG are very strong-coupled superconductors, one can conclude that these materials exhibit:

$$2\Delta(0) \gg 3.53 \times k_B T_c. \tag{11}$$

Thus, even in assumption that the Lande $g$-factor is unaltered (Eqs. 6,7) in MATBG and MATTG, Eq. 2 transforms in:

$$B_{c2,\parallel}(0) \lesssim B_{PCC}(0). \tag{12}$$

Based on this, the observed $B_{c2,\parallel}(0)$ enhancement in MATTG [1,2] agrees with the Pauli-Clogston-Chandrasekhar limiting field.

Apart from the problem of the Lande $g$-factor in atomically thin superconductors, here we present a model to deduce the ground-state superconducting energy gap, $\Delta(0)$, in these superconductors to, at least partially, clarify one multiplicative term in primary Eq. 6 of the Pauli-Clogston-Chandrasekhar limiting field.

The model is based on the analysis of both perpendicular, $B_{c2,\perp}(T)$ (when a magnetic field is applied in the direction perpendicular to the film surface), and parallel, $B_{c2,\parallel}(T)$, upper critical field data for atomically thin superconductors.

It should be noted that the scanning tunnelling spectroscopy (STS) is the most widely used technique to measure the superconducting energy gap, $\Delta(T)$, in atomically thin films [33]. However, the STS technique measures the out-of-plane component, $\Delta_c(T)$, of the gap, whereas while $B_{c2,\parallel}(0)$ in atomically thin superconductors depends on the in-plane, $\Delta_{ab}(0)$, gap component.

For anisotropic superconductors in-plane, $B_{c2,\perp}(T)$, and out-of-plane, $B_{c2,\parallel}(T)$, the upper critical fields within the Ginzburg-Landau theory are given by [34]:

$$B_{c2,perp}(T) = \frac{\phi_0}{2\pi} \times \frac{1}{\xi_{ab}^2(T)} \tag{13}$$

$$B_{c2,\parallel}(T) = \frac{\phi_0}{2\pi} \times \frac{1}{\xi_{ab}(T)} \times \frac{1}{\xi_c(T)} \tag{14}$$

where $\phi_0$ is the superconducting quantum flux. Recently, Farrar $et\ al$ [35] pointed out that the Bardeen-Cooper-Schrieffer (BCS) theory of superconductivity [36] considers two main Cooper pairs breaking mechanisms in atomically thin FeSe films under high magnetic field.

The first mechanism is the orbital pair breaking, which originates from the Lorentz force act on the charge of the paired electrons/holes. The respectful upper critical field for this mechanism is designated as the orbital limiting upper critical field, $B_{c2}^{orb}(T)$, and equations for these field are given by Eqs. 13,14 (where we omitted the use of superscript *orb* to simplify Eqs. 13, 14 and several following equations). The second mechanism is the pairs break by the spin-paramagnetic effect explained above (Eq. 6), which is designated as $B_{PCC}(T)$. Due to in many superconductors [37] (including atomically thin films [35]), $B_{c2}(T)$ is affected by both pair-breaking mechanisms, there is a useful parameter $\alpha_M = \sqrt{2} \frac{B_{c2}^{orb}(0)}{B_{PCC}(0)}$, introduced by Maki [38]. This parameter shows the relative strength between the orbital and the paramagnetic pair-breaking mechanisms for a given type-II superconductor. It should be noted, that Werthamer, Helfand, and Hohenberg developed a theory [39] which includes both pair-breaking mechanisms, accounted by the Maki parameter, $\alpha_M$, and the spin-orbit constant, $\lambda_{so}$ [39]. Both $\alpha_M$ and $\lambda_{so}$ have angular dependence and these parameters are utilized for the analysis of atomically thin superconductors (see, for instance, [35]).

Considering that in Bardeen-Cooper-Schrieffer (BCS) theory of superconductivity [36]:

$$\frac{1}{\xi(0)} = \frac{\pi}{\hbar} \times \frac{\Delta(0)}{v_F} \tag{15}$$

where $v_F$ is the Fermi velocity and the ground-state out-of-plane upper critical field is:

$$B_{c2,\parallel}(0) \propto \Delta_{ab}(0) \times \Delta_c(0) \tag{16}$$

It should be noted that even in bulk elemental fcc superconductors, for instance, in classical BCS superconductor aluminium [40-42], the superconducting energy gap exhibit crystallographic anisotropy, which is one of the important conclusions of the BCS theory. It should be mentioned detailed study of the superconducting gap anisotropy in aluminium reported by Blackford [42], who showed that pure bulk aluminium exhibits $\Delta_{min}(0) = 145 \, \mu eV$ and $\Delta_{max}(0) = 198 \, \mu eV$. This implies that in bulk samples $B_{PCC}(0)$, $B_{c2,\parallel}(0)$, $B_{c2,perp}(0)$, and $\frac{2\Delta(0)}{k_B T_c}$ can be different by factor of $\frac{\Delta_{max}(0)}{\Delta_{min}(0)} \cong 1.4$, depends on sample orientation and measuring technique.

Because Eqs. 14,16 have two independent multiplicative terms, it is impossible to fit $B_{c2,\parallel}(T)$ data to Eqs. 14,16 to keep both terms as free-fitting parameters. In other words, the utilization of Eq. 14 for $B_{c2,\parallel}(T)$ data analysis strictly depends on the chosen mathematical approximative function for temperature-dependent anisotropy, $\gamma_\xi(T)$:

$$\gamma_\xi(T) = \frac{\xi_{ab}(T)}{\xi_c(T)} = \frac{B_{c2,\parallel}(T)}{B_{c2,perp}(T)} \tag{17}$$

$$B_{c2,\parallel}(T) = \frac{\phi_0}{2\pi} \times \frac{1}{\xi_{ab}^2(T)} \times \gamma_\xi(T) \tag{18}$$

for given material.

Surprisingly, this problem does not exist in atomically thin superconductors, because it was first shown by Tinkham and co-workers [43-45], for conditions of very thin films:

$$d_{sc} \ll \xi(0) \tag{19}$$

$$d_{sc} \ll \lambda(0) \tag{20}$$

where $d_{sc}$ is the thickness of the superconducting layer, and Eq. 14 transforms in [43-45]:

$$B_{c2,\parallel}(T) = \frac{\phi_0}{2\pi} \times \frac{1}{\xi_{ab}(T)} \times \frac{\sqrt{12}}{d_{sc}}, \tag{21}$$

and, thus, one can conclude that Eq. 16 transforms in:

$$B_{c2,\parallel}(0) \propto \frac{\sqrt{12}}{d_{sc}} \times \Delta_{ab}(0), \tag{22}$$

As a result, $B_{c2,\parallel}(0)$ values described by Eqs. 21,22 can be orders of magnitude larger than $B_{c2,\parallel}(0)$ expected from Eq. 14, even if all fundamental parameters of the superconductor, such as $\xi_{ab}(0)$, $\xi_c(0)$, $\Delta_{ab}(0)$ and $\Delta_c(0)$, remain unchanged.

Truly, considering the anisotropy, $\gamma_\xi(0)$, for atomically thin film:

$$\gamma_\xi(0) = \frac{B_{c2,\parallel}(0)}{B_{c2,perp}(0)} = \sqrt{12} \times \left[ \frac{\xi_{ab}(0)}{d_{sc}} \right] \tag{23}$$

and typical values for $\xi_{ab}(0) \sim 50\ nm$ and $d_{sc} = 1\ nm$ for MATBG [31,46], we can calculate:

$$\gamma_\xi(0) = 200 \tag{24}$$

$$B_{c2,\parallel}(0) = 200 \times B_{c2,\perp}(0) \tag{25}$$

It should be noted that any Type-II superconductor has three fundamental fields, i.e. the lower critical field, $B_{c1}$, the upper critical field, $B_{c2}$, and the Pauli-Clogston-Chandrasekhar field, $B_{PCC}$. These fields are not independent from each other and they linked through primary superconducting parameters $\lambda$, $\xi$, $\Delta$, etc, of the material. Eq. 23 demonstrates that atomically thin superconductors have additional parameter which is the film thickness, $d_{sc}$. It should be stressed that Eq. 23 is not the only equation which utilizes the film thickness, $d_{sc}$, to rescale one of the fundamental values (in this case, the upper critical field) for thin film superconductors. The approach to utilize $d_{sc}$ as an additional fundamental constant for the case of very thin superconductors was recognized nearly six decades ago, when Pearl [47] showed that the characteristic length, $\lambda_{\perp,eff}(T)$, for the decay of screening currents of the vortex in thin film depends on the film thickness:

$$\lambda_{\perp,eff}(T) = \frac{2\lambda^2(T)}{d_{sc}} = \lambda(T) \times \frac{2\lambda(T)}{d_{sc}}. \tag{26}$$

Eq. 26 can be presented in the similar form to Eq. 23:

$$\gamma_\lambda(0) = \frac{\lambda_{\perp,eff}(0)}{\lambda(0)} = 2 \times \left[\frac{\lambda(0)}{d_{sc}}\right] \tag{27}$$

The difference between Eqs. 23 and 27 is the multiplicative prefactor $\sqrt{3}$. However, these equations demonstrate general approach that different fundamental superconducting values in atomically thin superconductors depend on the ratio of the bulk sample value divided by the film thickness, $d_{sc}$. We depicted the ratio by the square brackets in Eqs. 23 and 27.

In flavor of this approach, the geometrical enhancement factor (with the respectful ratio) should apply for two other fundamental fields in Type-II superconductors, i.e. for the lower critical field, $B_{c1}$, and for the the Pauli-Clogston-Chandrasekhar limiting field, $B_{PCC,\parallel}(0)$. Based on a fact that the Pauli-Clogston-Chandrasekhar limiting field in many cases is close (by its value) to the upper critical field, here we propose to apply the same geometrical enhancement factor for $B_{PCC,\parallel}(0)$ as for the $B_{c2,\parallel}(0)$ field:

$$\frac{B_{PCC,\parallel}(0)}{B_{PCC,perp}(0)} = \sqrt{12} \times \frac{\xi_{ab}(0)}{d_{sc}} \qquad (28)$$

and, thus, the primary Eq. 3 (even under the assumption of *s*-wave weak-coupling superconductivity) transforms into:

$$B_{PCC,\parallel}(0) = \sqrt{12} \times \frac{\xi_{ab}(0)}{d_{sc}} \times B_{PCC,\perp}(0) \cong 370 \times T_c \qquad (29)$$

which by 2 orders of magnitude exceeds observed in the experiment [1-17] enhancement in $B_{c2,\parallel}(0)$ (Eq. 2) for atomically thin superconductors.

It should be stressed that Eq. 29 originates from the GL theory [48] and the sample geometry (as it was first proposed by Tinkham and co-workers [43-45]), without any assumptions about the emergence of a new physical phenomenon, such as Ising superconductivity [1,24], in atomically thin superconductors.

Based on the above, one of the remaining tasks is to deduce $\Delta_{ab}(0)$ in atomically thin films and to compare this value with its counterpart in bulk materials.

One of the techniques available to achieve this was introduced by Talantsev *et al* [49,50], who proposed an equation for the self-field critical current density, $J_c(\text{sf},T)$, in type-II thin-film superconductors:

$$J_c(sf,T) = \frac{\phi_0}{4\pi\mu_0} \times \frac{ln(\kappa_c(T)) + 0.5}{\lambda_{ab}^3(T)}. \qquad (30)$$

where $\mu_0 = 4\pi \times 10^{-7} \ N/A^2$ is the permeability of free space, $\lambda_{ab}(T)$ is the in-plane London penetration depth, and $\kappa_c(T) = \frac{\lambda_{ab}(T)}{\xi_{ab}(T)}$ is temperature-dependent Ginzburg-Landau parameter. Later [51,52], equation (Eq. 30) has been extended to both type-I and type-II superconductors:

$$J_c(sf, T) = \frac{\phi_0}{4\pi\mu_0} \times \frac{ln\left(1+\sqrt{2}\times\kappa_c(T)\right)}{\lambda_{ab}^3(T)}.$$ (31)

Considering that $\lambda_{ab}(T)$ can be calculated from $\Delta_{ab}(T)$ (all expressions below are given for s-wave superconductors):

$$\lambda_{ab}(T) = \frac{\lambda_{ab}(0)}{\sqrt{1-\frac{1}{2k_BT}\int_0^\infty \frac{d\varepsilon}{\cosh^2\left(\frac{\sqrt{\varepsilon^2+\Delta_{ab}^2(T)}}{2k_BT}\right)}}},$$ (32)

where $k_B$ is the Boltzmann constant. The temperature-dependent superconducting gap can be expressed by the equation given by Gross et al [53]:

$$\Delta_{ab}(T) = \Delta_{ab}(0)\times\tanh\left[\frac{\pi k_B T_c}{\Delta_{ab}(0)} \times \sqrt{\eta\frac{\Delta C}{\gamma T_c}\left(\frac{T_c}{T}-1\right)}\right],$$ (33)

where $\frac{\Delta C}{\gamma T_c}$ is the relative jump in electronic specific heat at $T_c$ (where $\gamma$ is so-called Sommerfeld constant), and $\eta$ = 2/3 (for s-wave superconductors). Thus, primary parameters of the superconducting state of atomically thin superconductors:

1. Ground state superconducting energy gap, $\Delta_{ab}(0)$;

2. Relative jump in electronic specific heat at the transition temperature, $\frac{\Delta C}{\gamma T_c}$;

3. Ground state in-plane London penetration depth, $\lambda_{ab}(0)$;

4. Transition temperature, $T_c$;

5. Gap-to-transition temperature ratio, $\frac{2\Delta_{ab}(0)}{k_B T_c}$.

can be deduced from the fit of the measured $J_c(sf, T)$ to Eqs. 22-25. Thus, the superconducting parameters of thin films of Ga [50], FeSe [50], InN [54], TaS$_2$ [50], PdTe$_2$ [55], MoS$_2$ [5,50], α-Mo$_2$C [50], NbSe$_2$ [50,56], MATBG [57], and IrTe$_2$ [58] were deduced.

However, $J_c(sf, T)$ measurements in atomically thin superconductors are perhaps one of the most challenging experiments and, more commonly, magnetoresistance datasets, $R(T,B)$, are measured from which $B_{c2,perp}(T)$ and $B_{c2,\parallel}(T)$, or solely $B_{c2,\parallel}(T)$, can be extracted.

For these more common experiments, we introduced a model that allowed us to deduce the following:

1. Ground state superconducting energy gap, $\Delta_{ab}(0)$;

2. Relative jump in electronic specific heat at the transition temperature, $\frac{\Delta C}{\gamma T_c}$;

3. Ground state in-plane superconducting coherence length, $\xi_{ab}(0)$;

4. Transition temperature, $T_c$;

5. Gap-to-transition temperature ratio, $\frac{2\Delta_{ab}(0)}{k_B T_c}$.

and demonstrated the model applicability for atomically thin superconductors Al, Sn, NbSe$_2$, MoS$_2$, TTG, and WTe$_2$.

## 2. Model description

The model is based on two primary ideas:

1. Superconducting coherence length, $\xi_{ab}(T)$, is linked to the London penetration depth, $\lambda_{ab}(T)$, through the Ginzburg-Landau parameter, $\kappa_c(T) = \frac{\lambda_{ab}(T)}{\xi_{ab}(T)}$, for which we adopted temperature dependence proposed by Gor'kov [59]:

$$\kappa_c(T) = \kappa_c(0) \times \left(1 - 0.2429 \times \left(\frac{T}{T_c}\right)^2 + 0.0396 \times \left(\frac{T}{T_c}\right)^4\right) \qquad (34)$$

Based on Eq. 34, the temperature-dependent coherence length, $\xi_{ab}(T)$, can be represented as follows:

$$\xi_{ab}(T) = \frac{\xi_{ab}(0)}{\left(1 - 0.2429 \times \left(\frac{T}{T_c}\right)^2 + 0.0396 \times \left(\frac{T}{T_c}\right)^4\right)} \sqrt{\frac{1}{1 - \frac{1}{2k_B T} \int_0^\infty \frac{d\varepsilon}{\cosh^2\left(\frac{\sqrt{\varepsilon^2 + \Delta_{ab}^2(T)}}{2k_B T}\right)}}}, \qquad (35)$$

where temperature dependent superconducting energy gap, $\Delta_{ab}(T)$, is given by Eq. 34.

By substituting of Eq. 35 in Eqs. 13,21, one can obtain a system of two equations which describe the upper critical field in atomically thin films (i.e. films for which conditions of Eqs. 19,20 are satisfied):

$$
\begin{cases}
B_{c2,perp}(T) = \dfrac{\phi_0}{2\pi}\dfrac{1}{\xi_{ab}^2(T)} = \dfrac{\phi_0}{2\pi}\dfrac{\left(1 - 0.2429 \times \left(\frac{T}{T_c}\right)^2 + 0.0396 \times \left(\frac{T}{T_c}\right)^4\right)^2}{\xi_{ab}^2(0)}\left(1 - \dfrac{1}{2k_BT}\int\limits_{0}^{\infty}\dfrac{d\varepsilon}{\cosh^2\left(\dfrac{\sqrt{\varepsilon^2 + \Delta_{ab}^2(T)}}{2k_BT}\right)}\right) & (36) \\[4ex]
B_{c2,\parallel}(T) = \dfrac{\phi_0}{2\pi}\dfrac{1}{\xi_{ab}(T)}\dfrac{\sqrt{12}}{d_{sc}} = \dfrac{\phi_0}{2\pi}\dfrac{\left(1 - 0.2429 \times \left(\frac{T}{T_c}\right)^2 + 0.0396 \times \left(\frac{T}{T_c}\right)^4\right)}{\xi_{ab}(0)}\dfrac{\sqrt{12}}{d_{sc}}\sqrt{1 - \dfrac{1}{2k_BT}\int\limits_{0}^{\infty}\dfrac{d\varepsilon}{\cosh^2\left(\dfrac{\sqrt{\varepsilon^2 + \Delta_{ab}^2(T)}}{2k_BT}\right)}} & (37)
\end{cases}
$$

Eq. 36 was used to fit $B_{c2,\perp}(T)$ in MATBG [57] and thin film Nd$_{0.8}$Sr$_{0.2}$NiO$_2$ [60], while Eq. 37 is proposed herein.

2. The second novelty of this approach (Eqs. 36,37) is that the two datasets, $B_{c2,perp}(T)$ and $B_{c2,\parallel}(T)$, are simultaneously fitted to Eqs. 36,37. This type of fitting is also known as global data fit. As a result, a common set of free-fitting parameters, $\Delta_{ab}(0)$, $\frac{\Delta C}{C}$, $\xi_{ab}(0)$, $T_c$, $\frac{2\Delta(0)}{k_BT_c}$, and $d_{sc}$ can be deduced from the experimental $B_{c2,perp}(T)$ and $B_{c2,\parallel}(T)$ datasets. For cases, when only $B_{c2,\parallel}(T)$ datasets were reported, we used Eq. 37 with a fixed $d_{sc}$ to the reported layer thickness to extract the fundamental superconducting parameters for these films.

## 3. Results

### 3.1. Sn films

In 1962 Blumberg and Douglass [23,24] discovered a many-fold increase in $B_{c2,\parallel}(T)$ vs. the decrease in the superconductor thickness, $d_{sc}$, when they studied thin films of tin. In Fig.

1, we show $B_{c2,\parallel}(T)$ data and fit them to Eq. 37 for a Sn film with $d_{sc} = 36\ nm$ reported by Bloomberg [23]. Because the $B_{c2,\parallel}(T)$ dataset for this film was measured within a narrow temperature range $0.79 \leq \frac{T}{T_c} \leq 1.0$, to make it possible to deduce the ground state energy gap, $\Delta(0)$, for this film, we fixed $\frac{\Delta C}{C}$ to its reported value of 1.68 [28] in Eq. 37. This approach is based on the similarity proposed by Talantsev *et al* [46], who reduced the number of free-fitting parameters in Eqs. 29-31 by fixing $\frac{\Delta C}{C}$ for $J_c(sf, T)$ datasets, when fitted datasets do not contain enough raw data points to deduce $\Delta(0)$ with the required accuracy.

In the result, deduced $\xi(0) = 61.8 \pm 1.1\ nm$ satisfies the condition of $d_{sc} < \xi(0)$ and deduced $\frac{2\Delta(0)}{k_B T_c} = 3.7 \pm 0.3$ is in excellent agreement with the value of $\frac{2\Delta(0)}{k_B T_c} = 3.705$ reported by Carbotte [29] for electron-phonon mediated superconductivity in bulk tin.

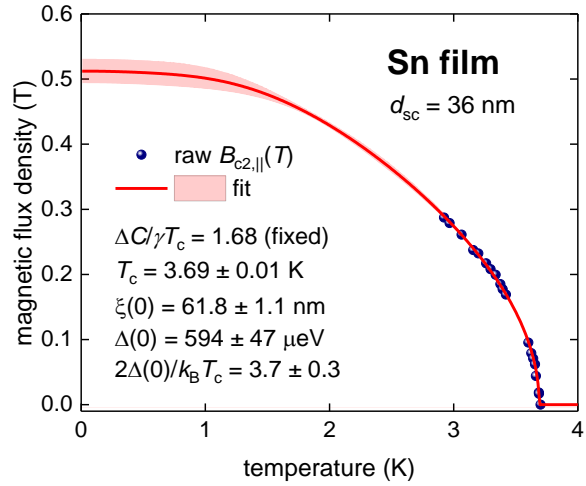

**Figure 1.** $B_{c2,\parallel}(T)$ data and fit to Eq. 37 for a Sn film with thickness $d_{sc} = 36\ nm$ (raw data reported by Bloomberg [22]). $\frac{\Delta C}{\gamma T_c} = 1.68$ was fixed at the value reported by Carbotte [29]. Deduced parameters $T_c = 3.69 \pm 0.01\ K$, $\xi(0) = 61.8 \pm 1.1\ nm$, $\Delta(0) = 594 \pm 47\ \mu eV$, $\frac{2\Delta(0)}{k_B T_c} = 3.7 \pm 0.3$. The goodness of fit is 0.9990. The 95% confidence bands are shown by the shaded areas.

To demonstrate the global fit to Eqs. 36,37 works for thin films of tin in Figure 2 we show experimental $B_{c2,perp}(T)$ and $B_{c2,\parallel}(T)$ datasets for sample TD3 ($d_{sc} = 13.6\ nm$) reported by Harper and Tinkham [45]. For this fit (Fig. 2), we also fixed $\frac{\Delta C}{\gamma T_c} = 1.68$, whereas the other parameters were free. All deduced values (for instance, $\xi(0) = 70.6 \pm 1.5 \, nm$, $\Delta(0) = 613 \pm 58 \, \mu eV$, and $\frac{2\Delta(0)}{k_B T_c} = 4.0 \pm 0.4$) are in a remarkable agreement with values deduced for data reported by Bloomberg (Fig. 1 and Ref. 23), and the values calculated by Carbotte [29] for bulk materials. For instance, the extracted $\Delta(0)$ for Bloomberg's data is $\Delta(0) = 594 \pm 47 \, \mu eV$ (Fig. 1) vs. the extracted $\Delta(0)$ for Harper's and Tinkham's data: $\Delta(0) = 613 \pm 58 \, \mu eV$ (Fig. 2). It should be stressed that this consistency is found for experimental data reported by two different groups with time intervals of six years between reports.

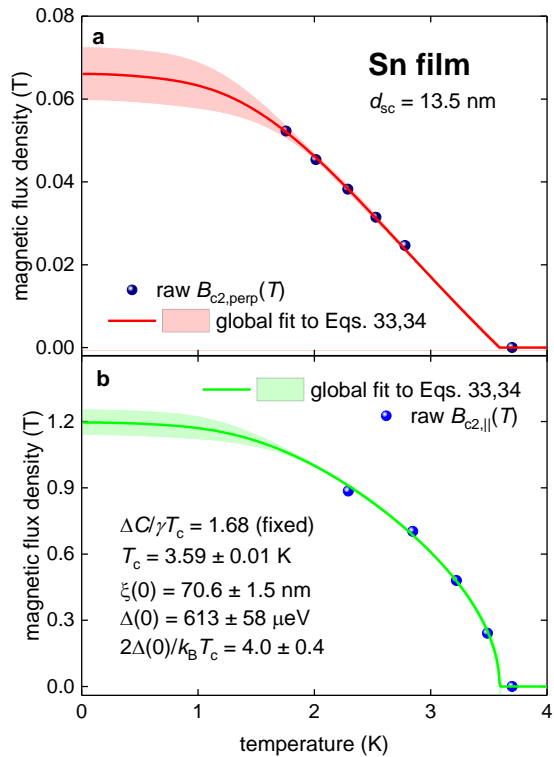

**Figure 2.** $B_{c2,perp}(T)$ (a) and $B_{c2,\parallel}(T)$ (b) datasets and global fits to Eqs. 36,37 for Sn film having thickness $d_{sc} = 13.5 \, nm$ (raw data reported by Harper and Tinkham [45]). $\frac{\Delta C}{\gamma T_c} = 1.68$ was fixed to the value reported by Carbotte [29]. Deduced parameters $T_c = 3.59 \pm 0.01 \, K$, $\xi(0) = 70.6 \pm 1.5 \, nm$, $\Delta(0) = 613 \pm 58 \, \mu eV$, $\frac{2\Delta(0)}{k_B T_c} = 4.0 \pm 0.4$. The goodness of fit is: (a) 0.9993 and (b) 0.9980. The 95% confidence bands are shown by the shaded areas.

## 3.2. Al films

A few-nanometre-thick aluminium film represents, perhaps, the most studied object of elemental metallic superconductors, owing to their prominent dependence of $T_c$, $\Delta(0)$, and $B_{c2,\parallel}(0)$ on the film thickness, $d_{sc}$ [25-27,61,62]. The highest $T_c = 5.7$ K was reported by Townsend *et al* [61] for aluminum film with $d_{sc} = 2\ nm$, while bulk aluminum exhibits $T_c = 1.12$ K [29].

In Figure 3, we show $B_{c2,\parallel}(T)$ dataset reported for Al film with $d_{sc} = 5\ nm$ by Tedrow and Meservey [63] and the data fit to Eq. 35. The deduced $\Delta(0), \frac{\Delta C}{\gamma T_c}$, and $\frac{2\Delta(0)}{k_B T_c}$ values showed that this Al film is strongly coupled superconductor. This result agrees with the general trend [25-27,61-63], where the thinning of Al causes a significant enhancement in $T_c$, $\Delta(0)$, and $B_{c2,\parallel}(0)$.

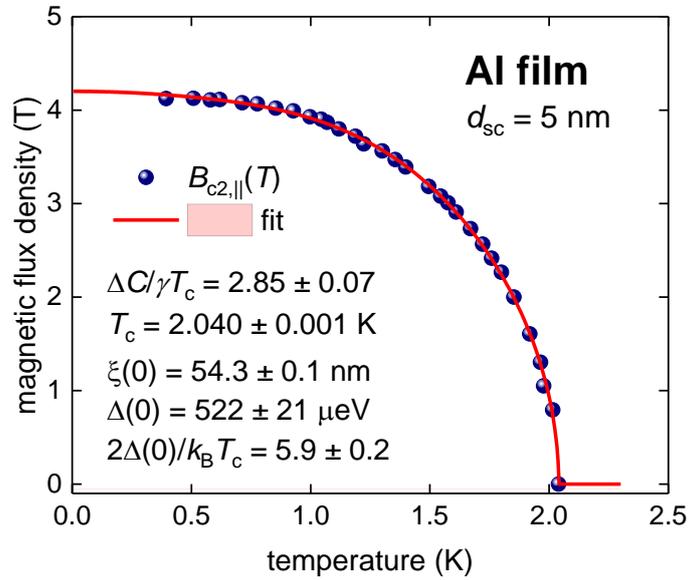

**Figure 3.** $B_{c2,\parallel}(T)$ data and fit to Eq. 37 for an Al film with thickness $d_{sc} = 5\ nm$ (raw data reported by Tedrow and Meservey [63]). Deduced parameters $T_c = 2.040 \pm 0.001\ K$, $\xi(0) = 54.3 \pm 0.1\ nm$, $\Delta(0) = 522 \pm 21\ \mu eV$, $\frac{\Delta C}{\gamma T_c} = 2.85 \pm 0.07$, $\frac{2\Delta(0)}{k_B T_c} = 5.9 \pm 0.2$. The goodness-of-fit was 0.9994. The 95% confidence bands are indicated by the shaded areas.

It should be stressed, that if for Pauli-Clogston-Chandrasekhar limiting field calculation, one uses deduced $T_c = 2.04\ K$ (Fig. 3) and weak-coupling Eq. 3, then:

$$B_{PCC,weak-coupling}(0) = 1.86 \times T_c = 3.79\ T < B_{c2,\parallel}(0) = 4.2\ T. \tag{38}$$

This means that in this Al film, the Pauli-Clogston-Chandrasekhar limiting field was violated. However, if proper Eq. 6 and deduced $\Delta(0) = 522\ \mu eV$ is used for calculations, then:

$$B_{PCC}(0) = \frac{\Delta(0)}{\sqrt{g} \times \mu_B} = 6.4\ T > B_{c2,\parallel}(0) = 4.2\ T. \tag{39}$$

Thus, this Al film complies with Pauli-Clogston-Chandrasekhar limiting field, because $B_{c2,\parallel}(0)$ is significantly lower than $B_{CPP}(0)$.

Practically identical values for $\Delta(0)$, $\frac{\Delta C}{\gamma T_c}$, and $\frac{2\Delta(0)}{k_B T_c}$ were deduced for the 4 nm thick Al film, for which raw $B_{c2,\parallel}(0)$ data were reported by Tedrow and Meservey in their Figure 7 [27]. $B_{c2,\parallel}(0)$ data and fit for the Al film are shown in Figure 4. Calculated $B_{CPP}(0)$, based on Eq.6, is well above extrapolated $B_{c2,\parallel}(0)$:

$$B_{PCC}(0) = \frac{\Delta(0)}{\sqrt{g} \times \mu_B} = 6.3\ T > B_{c2,\parallel}(0) = 4.2\ T. \tag{40}$$

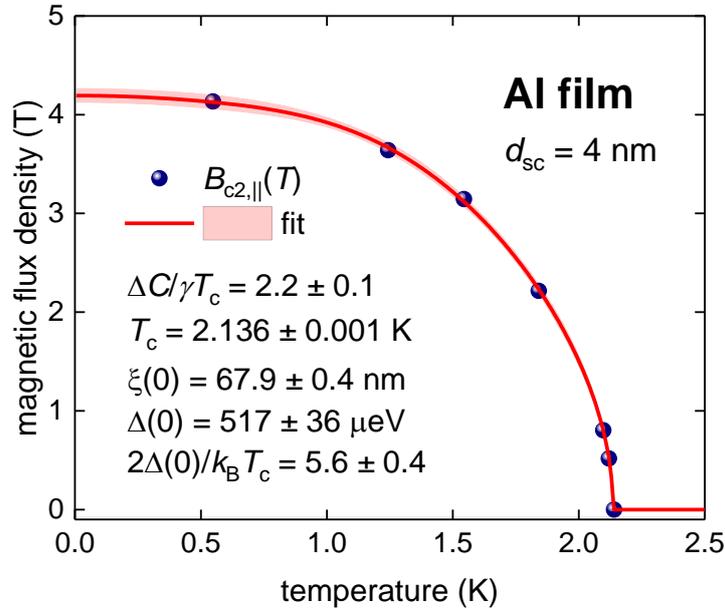

**Figure 4.** $B_{c2,\parallel}(T)$ data and fit to Eq. 37 for Al film having thickness $d_{sc} = 4\ nm$ (raw data reported by Tedrow and Meservey [27]). Deduced parameters $T_c = 2.136 \pm 0.001\ K$, $\xi(0) = 67.9 \pm 0.4\ nm$, $\Delta(0) = 517 \pm 36\ \mu eV$, $\frac{\Delta C}{\gamma T_c} = 2.2 \pm 0.1$, $\frac{2\Delta(0)}{k_B T_c} = 5.6 \pm 0.4$. The goodness-of-fit is 0.9998. 95% confidence bands are shown by shaded areas.

### 3.3. Al film overcoated with Pt

Tedrow and Meservey [27] reported three $B_{c2,\parallel}(0)$ datasets for 4-nm thick Al films overcoated with thin Pt layers of different thicknesses, $d_{Pt}$. From the three reported $B_{c2,\parallel}(0)$ datasets, only one dataset with $d_{Pt} = 0.1\ nm$ has sufficient raw $B_{c2,\parallel}(0)$ data points to be fitted to Eq. 37 (Fig. 5). The deduced $\Delta(0) = 451 \pm 50\ \mu eV$ is in close proximity to the values deduced for the other Al films (Figs. 3,4). Calculated Pauli-Clogston-Chandrasekhar limiting field:

$$B_{PCC,perp}(0) = \frac{\Delta(0)}{\sqrt{g} \times \mu_B} = 5.5\ T < B_{c2,\parallel}(0) = 7.8\ T. \tag{41}$$

appeared to be lower than that of the extrapolated $B_{c2,\parallel}(0)$. One of the two possible interpretations of this result is that Pt overcoating can supress the Lande $g$-factor. However, it is much more natural to explain this result based on the geometrical field enhancement factor (Eqs. 23,28):

$$B_{PCC,\parallel}(0) = \sqrt{12} \times \frac{\xi_{ab}(0)}{d_{sc}} \times B_{PCC,perp}(0) = 31.5 \times 5.5\ T = 173\ T \tag{42}$$

which by 22 times exceeds extrapolated $B_{c2,\parallel}(0) = 7.8\ T$ (Fig. 5).

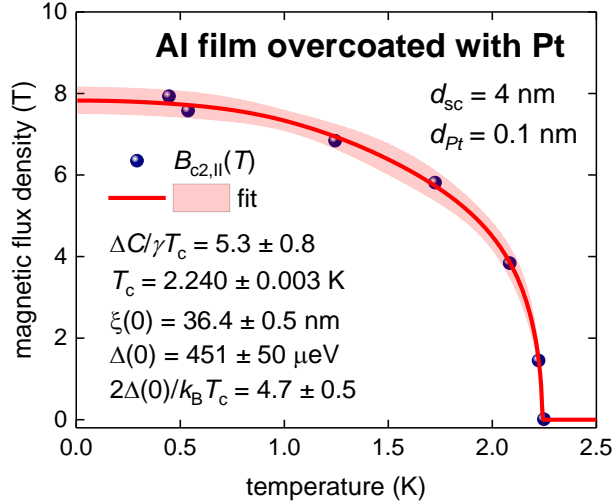

**Figure 5.** $B_{c2,\parallel}(T)$ data, and fit to Eq. 37 for Al film having thickness $d_{sc} = 4\ nm$ overcoated with Pt layer with thickness $d_{Pt} = 0.1\ nm$ (raw data reported by Tedrow and Meservey [27]). Deduced parameters $T_c = 2.240 \pm 0.003\ K$, $\xi(0) = 36.4 \pm 0.5\ nm$, $\Delta(0) = 451 \pm 50\ \mu eV$, $\frac{\Delta C}{\gamma T_c} = 5.3 \pm 0.8$, $\frac{2\Delta(0)}{k_B T_c} = 4.7 \pm 0.5$. The goodness-of-fit was 0.9988. The 95% confidence bands are indicated by the shaded areas.

### 3.4. Electric-field gated WTe₂ monolayer

Sajadi *et al* [27] reported $B_{c2,perp}(T)$ and $B_{c2,\parallel}(T)$ datasets for the WTe₂ monolayer sample in their Figures 2(b,c). Although these datasets were measured at slightly different doping states (i.e., $n_e = 2.0 \times 10^{13} \ cm^{-2}$ for $B_{c2,perp}(T)$ and $n_e = 1.8 \times 10^{13} \ cm^{-2}$ for $B_{c2,\parallel}(T)$), we performed a global fit for these datasets to Eqs. 36,37 in Fig. 6. By performing this fit it was assumed that the superconducting layer thickness, $d_{sc}$, is a free fitting parameter, because it is difficult to establish this parameter in electric-field-gated sample.

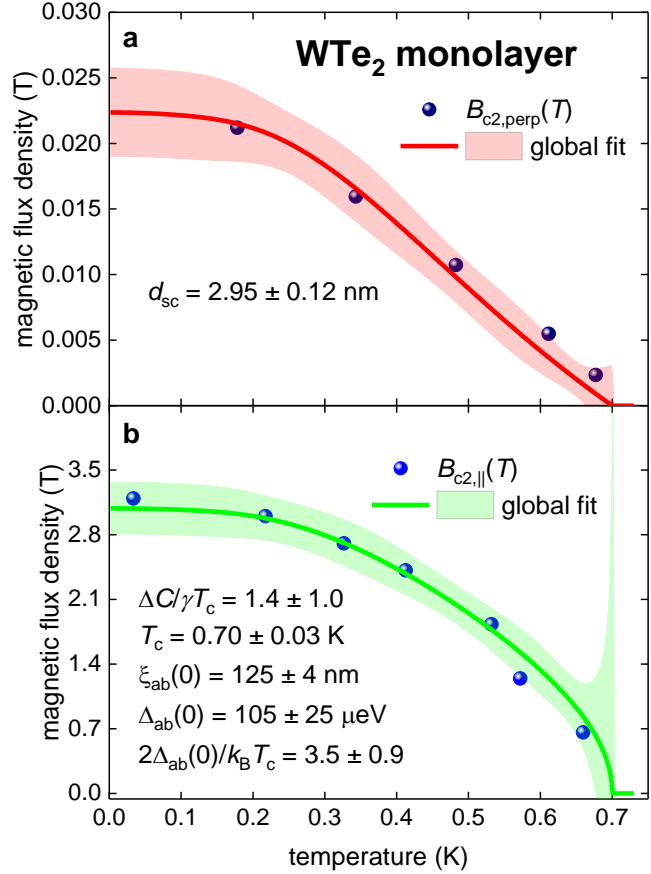

**Figure 6.** (a) $B_{c2,perp}(T)$ and (b) $B_{c2,\parallel}(T)$ data and global fit to Eqs. 36,37 for WTe₂ monolayer doped at (a) $n_e = 2.0 \times 10^{13} \ cm^{-2}$ and (b) $n_e = 1.8.0 \times 10^{13} \ cm^{-2}$ (raw data reported by Sajadi *et al* [27]). Deduced parameters are: $d_{sc} = 2.95 \pm 0.12 \ nm$, $T_c = 0.70 \pm 0.03 \ K$, $\xi_{ab}(0) = 125 \pm 4 \ nm$, $\Delta_{ab}(0) = 105 \pm 25 \ \mu eV$, $\frac{2\Delta_{ab}(0)}{k_B T_c} = 3.5 \pm 0.9$, $\frac{\Delta C}{\gamma T_c} = 1.4 \pm 1.0$. The goodness-of-fits are (a) 0.9718 and (b) 0.9764. The 95% confidence bands are indicated by the shaded areas.

All the deduced parameters agree well with the weak-coupling superconductivity in this electric-field-gated monolayer superconductor. Calculated Pauli-Clogston-Chandrasekhar limiting field:

$$B_{PCC,perp}(0) = \frac{\Delta(0)}{\sqrt{g} \times \mu_B} = 1.3 \ T < B_{c2,\parallel}(0) = 3.1 \ T. \tag{43}$$

appeared to be lower than that of the extrapolated $B_{c2,\parallel}(0)$. However, similarly to the case of the Al film overcoated with Pt (Fig. 5), this result can be explained based on the geometrical enhancement factor (Eqs. 23,28):

$$B_{PCC,\parallel}(0) = \sqrt{12} \times \frac{\xi_{ab}(0)}{d_{sc}} \times B_{PCC,perp}(0) = 31.5 \times 5.5 \ T = 55 \ T \qquad (44)$$

which by 18 times exceeds extrapolated $B_{c2,\parallel}(0) = 3.1 \ T$ (Fig. 6). Thus, this superconductor complies with Pauli-Clogston-Chandrasekhar limiting field.

### 3.5. Electric-field gated MoS₂ monolayer

Saito *et al* [4] reported $R_{sheet}(T,B)$ curves for a 20-nm thick MoS₂ single crystal in which the superconducting state was induced by electric-field gating. From the reported $R_{sheet}(T,B)$ curves measured at gate voltage of 5.5 V (the original curves are shown in Figs. 3(a,b) in Ref. 4) by utilizing 50% of the normal state resistance, we deduced $B_{c2,perp}(T)$ and $B_{c2,\parallel}(T)$ datasets, which are shown in Fig. 7 together with the global data fit to Eqs. 36,37. By performing the fit, it was assumed that the superconducting layer thickness, $d_{sc}$, is a free fitting parameter.

Deduced $\frac{2\Delta_{ab}(0)}{k_B T_c} = 8.1 \pm 1.4$ (at $V_{gate} = 5.5 \ V$, Fig. 7) is high, however we should note that Talantsev *et al* [46] reported $\frac{2\Delta_{ab}(0)}{k_B T_c} = 6.1 \pm 1.5$ for electric-field gated MoS₂ (with $V_{gate} = 2.2 \ V$) which was extracted from the analysis of $J_c(\text{sf},T)$ measured by Costanzo *et al* [64]. Thus, deduced by us the ratio of $\frac{2\Delta_{ab}(0)}{k_B T_c} = 8.1 \pm 1.4$ (at $V_{gate} = 5.5 \ V$) is in the same ballpark which is additional confirmation of strong-coupled superconductivity in electric-field gated MoS₂.

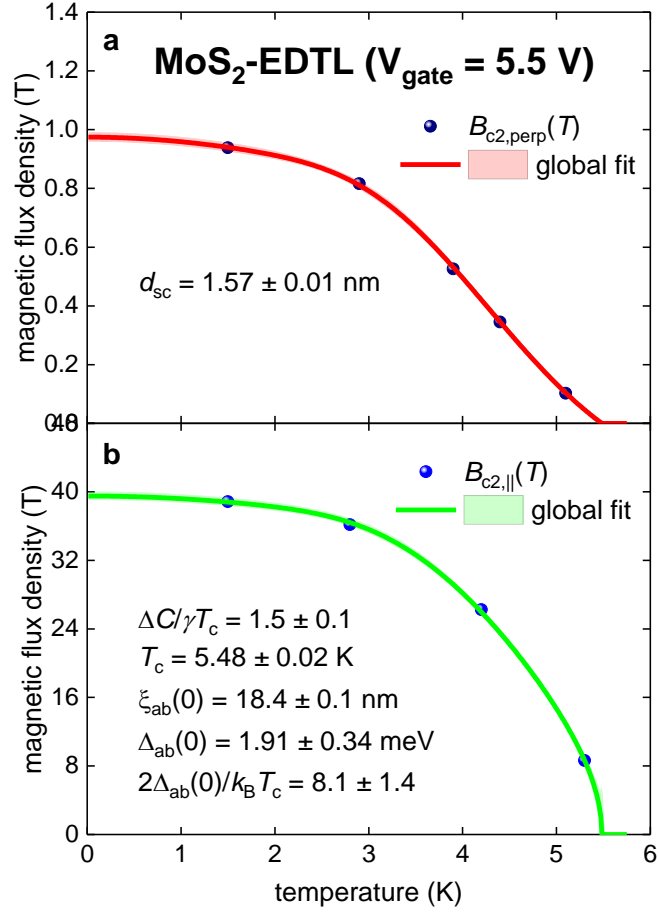

**Figure 7.** (a) $B_{c2,perp}(T)$ and (b) $B_{c2,\parallel}(T)$ data and global fit to Eqs. 36,37 for MoS$_2$-EDTL (raw $R_{\text{sheet}}(T,B)$ curves reported by Saito *et al* [4]). Deduced parameters are: $d_{sc} = 1.57 \pm 0.01\ nm$, $T_c = 5.48 \pm 0.02\ K$, $\xi_{ab}(0) = 18.4 \pm 0.1\ nm$, $\Delta_{ab}(0) = 1.91 \pm 0.34\ meV$, $\frac{2\Delta_{ab}(0)}{k_B T_c} = 8.1 \pm 1.4$, $\frac{\Delta C}{\gamma T_c} = 1.5 \pm 0.1$. The goodness of fits was (a) 0.9998 and (b) 0.9997. The 95% confidence bands are indicated by the shaded areas.

Calculated Pauli-Clogston-Chandrasekhar limiting field:

$$B_{PCC,perp}(0) = \frac{\Delta(0)}{\sqrt{g} \times \mu_B} = 23.3\ T < B_{c2,\parallel}(0) = 39.5\ T. \qquad (45)$$

is lower than extrapolated $B_{c2,\parallel}(0)$. However, this result again can be explained based on geometrical enhancement factor (Eqs. 23,28):

$$B_{PCC,\parallel}(0) = \sqrt{12} \times \frac{\xi_{ab}(0)}{d_{sc}} \times B_{PCC,perp}(0) = 946\ T. \qquad (46)$$

The magnetic flux density, $B_{PCC,\parallel}(0)$, indicated in Eq. 46 was created only several times on Earth during extraordinary pulsed magnetic field experiments [65-67]. Thus, observed in experiment $B_{c2,\parallel}(T \to 0\ K) = 39.5\ T$ is still by more than one order of magnitude lower than the Pauli-Clogston-Chandrasekhar limiting field.

### 3.6. NbSe₂ bilayer

de la Barrera *et al* [7] reported $B_{c2,perp}(T)$ and $B_{c2,\parallel}(T)$ datasets for NbSe₂ bilayer sample in their Figures 3(e) [7]. We performed a global fit for these datasets using Eqs. 36,37 in Fig. 8. By performing this fit, it was assumed that the superconducting layer thickness, $d_{sc}$, is a free fitting parameter.

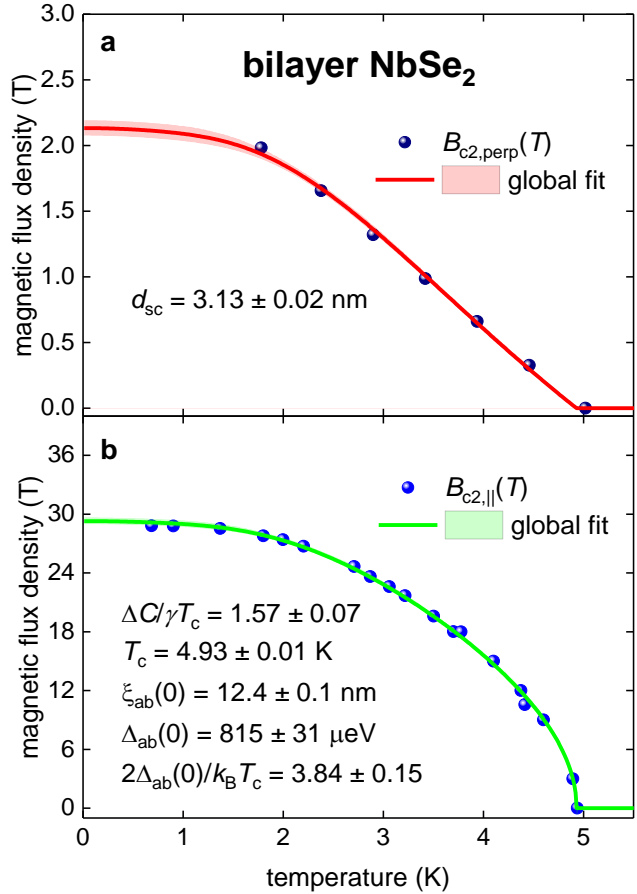

**Figure 8.** (a) $B_{c2,perp}(T)$ and (b) $B_{c2,\parallel}(T)$ data and global fit to Eqs. 36,37 for bilayer NbSe₂ (raw data by de la Barrera *et al* [7]). Deduced parameters are: $d_{sc} = 3.13 \pm 0.02\ nm$, $T_c = 4.93 \pm 0.01\ K$, $\xi_{ab}(0) = 12.4 \pm 0.1\ nm$, $\Delta_{ab}(0) = 815 \pm 31\ \mu eV$, $\frac{2\Delta_{ab}(0)}{k_B T_c} = 3.84 \pm 0.15$, $\frac{\Delta C}{\gamma T_c} = 1.57 \pm 0.07$. The goodness of fits was (a) 0.9981 and (b) 0.9989. The 95% confidence bands are indicated by the shaded areas.

The deduced values for the primary superconducting parameters, that are $\Delta_{ab}(0), \frac{\Delta C}{\gamma T_c}$, and $\frac{2\Delta_{ab}(0)}{k_B T_c}$ are within the expected values for moderately strong-coupled superconductors. It should be mentioned that Figure 8 demonstrates the importance of large number of experimental upper critical field data points which cover, as wide as it is possible, full temperature range $0.0 < \frac{T}{T_c} \leq 1.0$. For this type of dataset, the global fit to Eqs. 36,37, it is possible to deduce $d_{sc}, T_c, \xi_{ab}(0), \Delta_{ab}(0), \frac{\Delta C}{\gamma T_c}, \frac{2\Delta_{ab}(0)}{k_B T_c}$, in atomically thin superconductors with high accuracy.

Calculated Pauli-Clogston-Chandrasekhar limiting field:

$$B_{PCC,perp}(0) = \frac{\Delta(0)}{\sqrt{g} \times \mu_B} = 10\ T < B_{c2,\|}(0) = 29\ T. \qquad (47)$$

is lower than the extrapolated $B_{c2,\|}(0)$. However, this result can be explained based on the geometrical enhancement factor (Eqs. 23,28):

$$B_{PCC,\|}(0) = \sqrt{12} \times \frac{\xi_{ab}(0)}{d_{sc}} \times B_{PCC,perp}(0) = 137\ T. \qquad (48)$$

### 3.7. Magic-angle twisted trilayer graphene

Finally, we present an analysis of several magic-angle twisted trilayer graphene (MATTG) samples for which raw $B_{c2,\|}(T)$ datasets were recently reported by Cao *et al* [1]. For our analysis, we used $B_{c2,\|}(T)$ datasets, which were deduced from experimental $R(T)$ curves by applying 10% of the normal state resistance criterion [1]. The importance of utilizing as low the possible resistive criterion for $B_{c2}(T)$ definition with following analysis by Eqs. 36,37 is based on the fact that one of the primary parameters which is deduced from the fit $\frac{\Delta C}{C}$ can be accurately measured only when

$$\frac{R(T)}{R_{norm}} \to 0, \qquad (49)$$

and this was clearly shown in recent precise experiments by Hirai *et al* [68]. In addition, Shang *et al* [69] in their precise experiments showed that the diamagnetic response from NbReSi superconductor was observed at the condition described by Eq. 49.

Another issue is that only $B_{c2,\parallel}(T)$ datasets were reported for MATTG [1,2]. Based on this, we fixed the superconducting layer thickness $d_{sc}$ in Eq. 36 to the assumed MATTG thickness of $d_{sc} = 1\ nm$. MATTG devices for which $B_{c2,\parallel}(T)$ datasets are analyzed herein have adjacent layers, which are sequentially twisted by $\theta$ and $-\theta$ angles, where $\theta \approx 1.57°$ and $1.44°$ (details can be found elsewhere [1,32]).

In Figure 9 we show four $B_{c2,\parallel}(T)$ datasets chosen to cover the full range of observed $T_c$ values. As shown in Fig. 9, all deduced $\frac{2\Delta_{ab}(0)}{k_B T_c}$ ratios are within the well-established range of this parameter variation for classical low-$T_c$ electron-phonon mediated superconductors [29]:

$$3.5 \leq \frac{2\Delta_{ab}(0)}{k_B T_c} \leq 4.7 \qquad (50)$$

Thus, there is the first concern within regard to the current interpretation [1,2,32] of MATTG as a strong-coupling superconductor because all deduced $\frac{2\Delta_{ab}(0)}{k_B T_c}$ values demonstrated that MATTG exhibits much lower $\frac{2\Delta_{ab}(0)}{k_B T_c}$ values in comparison with iron-based superconductors [30] and superhydride superconductors [70].

In addition, as reported by Cao *et al* [1], the effect of reentering the superconducting state in MATTG at high magnetic fields has been observed in many superconductors [71], including single-layer graphene with deposited tin nanodisks [72]. Thus, the effect of reentering of the superconducting state is also not a unique MATTG property.

The remaining task to show that MATTG (and presumably MATBG) does not demonstrate any unique superconducting properties that were not observed in other superconducting materials yet is to calculate $B_{PCC,\parallel}(0)$ and to show that these fields are well above the observed $B_{c2,\parallel}(0)$ values. These calculations are presented in Table I, from which it is clear that all MATTG samples exhibit an inequality:

$$B_{c2,\parallel}(0) \ll B_{PCC,\parallel}(0) \qquad (49)$$

because of the very large values of the $\sqrt{12} \times \frac{\xi_{ab}(0)}{d_{sc}}$ ratio, similarly to all other atomically thin superconductor.

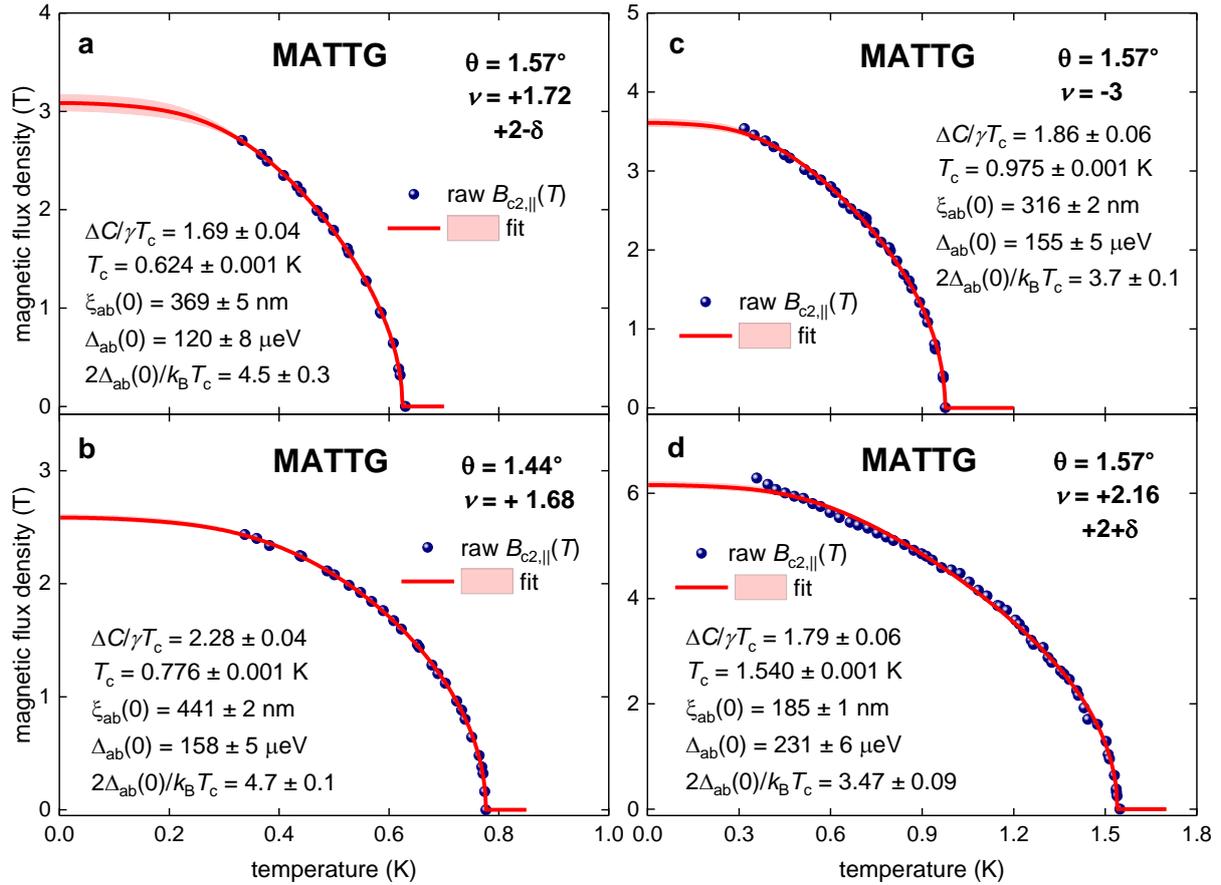

**Figure 9.** $B_{c2,\parallel}(T)$ datasets for the MATTG samples and data fits to Eq. 37 (raw data reported by Cao *et al* [1]). Assumed film thickness $d_{sc} = 1.0 \; nm$ (fixed). The deduced parameters are in Table I. The goodness of fits are: (a) 0.9998; (b) 0.9997; (c) 0.9987; (d) 0.9982. The 95% confidence bands are indicated by shaded areas.

**Table I.** Primary deduced parameters for MATTG samples showed in Figure 9.

| Sample designation in Figure 9 and in Ref. 1 | $T_c$ (K) | $\Delta_{ab}(0)$ ($\mu$eV) | $\frac{2\Delta_{ab}(0)}{k_B T_c}$ | $\frac{\Delta C}{\gamma T_c}$ | $B_{c2,\parallel}(0)$ (T) | $B_{PCC,perp}(0)$ (T) | $B_{PCC,\parallel}(0)$ (T) |
|---|---|---|---|---|---|---|---|
| **Panel a** Figure S1a $\theta = 1.57°$ $\upsilon = +1.72$ | 0.624 ± 0.001 | 120 ± 8 | 4.5 ± 0.3 | 1.69 ± 0.04 | 3.1 | 1.5 | 1917 |
| **Panel b** Figure S2b $\theta = 1.44°$ $\upsilon = +1.68$ | 0.776 ± 0.001 | 158 ± 5 | 4.7 ± 0.1 | 2.28 ± 0.04 | 2.6 | 1.9 | 2903 |
| **Panel c** Figure S2b $\theta = 1.57°$ $\upsilon = -3$ | 0.975 ± 0.001 | 155 ± 5 | 3.7 ± 0.1 | 1.86 ± 0.06 | 3.6 | 1.9 | 2080 |
| **Panel d** Figure S1b $\theta = 1.57°$ $\upsilon = +2.16$ | 1.540 ± 0.001 | 231 ± 6 | 3.47 ± 0.09 | 1.79 ± 0.06 | 6.2 | 2.8 | 1794 |

## 4. Discussion

It is interesting to mention that it is very likely that the first experimental observation of the emerge of the superconductivity in twisted atomically thin nanosheets was reported by Pan *et al* [73]. This research group prepared bulk $TaS_2$ samples by randomly restacked chemically exfoliated $TaS_2$ monolayers. In a result, randomly restacked bulk samples exhibit $T_c \sim 3$ K, compared with $T_c = 0.8$ K observed in ordered bulk 2H-$TaS_2$ compound. The randomness of the restacked samples was confirmed by the analysis of angular dependence of the upper critical field [74]. One of the most natural explanation of such $T_c$ enhancement is based on idea that in billions of randomly restacked $TaS_2$ monolayers, some neighbour monolayers are restacked at either magic angle twist [75], or at rhombohedral configuration [2]. Thus, the superconducting state in bulk randomly restacked $TaS_2$ samples is emerged similarly to bi- and tri-layer graphene. It is important to note, that Ma *et al* [76] reported that extrapolated $B_{c2,\parallel}(0) = 9.5\ T$ is significantly exceeded $B_{PCC}(0) = 5.8\ T$ (calculated by Eqs. 6-8) in bulk samples of randomly restacked $TaS_2$ monolayers. This is not a surprize, if we assume that the superconductivity in restacked TaS$_2$ originates from the magic angle or the rhombohedral twisted bi- or tri-layers TaS$_2$.

There is a need to stress that if even MATTG and other atomically thin superconductors exhibited a violation of the Pauli-Clogston-Chandrasekhar limiting field, it is still not any extraordinary property of these materials because many bulk superconductors exhibit this violation.

Truly, it can be mentioned the observation of the violation in bulk Ta$_2$Pd$_x$S$_5$ (x = 0.92) samples reported by Lu *et al.* [77], who calculated $B_{PCC}(0) \cong 10\ T$, based on Eqs. 6-8 and measured $T_c \sim 6\ K$, while measured $B_{c2,||b}(T = 0.50\ K) \cong 30\ T$ and $B_{c2,\perp b}(T = 0.55\ K) \cong 10\ T$, where *b* is the crystallographic axis.

There are several ceramic compounds for which the relation of $B_{c2}(T \to 0\ K) > B_{PCC}(0)$ was reported. For instance we can mentioned cubic centrosymmetric $\eta$-carbide Nb$_4$Rh$_2$C$_{1-\delta}$ [78] for which measured upper critical field $B_{c2}(T \to 0\ K) > 28\ T$ exceeds $B_{PCC}(0) = 18.1\ T$ (calculated in the assumption of weak-coupled *s*-wave superconductivity, i.e. based on Eqs. 6-8 and measured $T_c = 9.75\ K$). Ma *et al.* [79] reported that $B_{c2}(T \to 0\ K) > B_{PCC}(0)$ in Ti$_4$Rh$_2$O and Ti$_4$Ir$_2$O ceramics. Górnicka *et al.* [80], reported the relation of $B_{c2}(T \to 0\ K) > B_{PCC}(0)$ for NbIr$_2$B$_2$ and TaIr$_2$B$_2$ ceramics.

Thus, it can be concluded that the relation of $B_{c2}(T \to 0\ K) > B_{PCC}(0)$ (where the latter is calculated by Eq. 6-8) observes in many bulk ceramic materials, and this observation cannot be designated as unique extraordinal property of twisted graphene superlattices.

## 4. Conclusions

A new model for deducing several primary superconducting parameters in atomically thin superconductors, that is in-plane ground state superconducting energy gap, $\Delta_{ab}(0)$, relative jump in electronic specific heat at the transition temperature, $\frac{\Delta C}{\gamma T_c}$, in-plane ground state superconducting coherence length, $\xi_{ab}(0)$, transition temperature, $T_c$, and gap-to-transition temperature ratio, $\frac{2\Delta_{ab}(0)}{k_B T_c}$, from the upper critical field data was proposed. The analysis was performed for a few atomic layers of thick Al, Sn, NbSe$_2$, MoS$_2$, MATTG, and WTe$_2$are and showed that all the films except electric-field-gated MoS$_2$ exhibit moderately strong-coupling pairing.

The observed enhancement of parallel upper critical field, $B_{c2,\parallel}(0)$, in atomically thin superconductors, which is for some samples exceeded the Pauli-Clogston-Chandrasekhar limiting field calculated for perpendicular to the film surface direction of applied magnetic field, $B_{CPP,perp}(0)$, is explained based on geometrical enhancement factor which arises solely from small samples thickness. This enhancement is not related to the appearance of any new exotic pairing mechanism or new extraordinary physical phenomena, at least in all materials to date.

**Competing interests**

The authors declare no competing interests.

**Data availability statement**

The data that support the findings of this study are available upon reasonable request from the authors.

**Acknowledgements**

The author acknowledges financial support provided by the Ministry of Science and Higher Education of Russia (theme "Pressure" No. AAAA-A18-118020190104-3).